\documentclass {aa}
\usepackage{times}
\usepackage{graphics}
\usepackage{epsfig}

\makeatletter
\@ifundefined{chapter}{\def\thebibliography#1{
\section*{References}                          
  \list
  {\relax}{\setlength{\labelsep}{0em}
        \setlength{\itemindent}{-\bibhang}
        \setlength{\itemsep}{\parskip}          
        \setlength{\parsep}{0pt}
        \setlength{\leftmargin}{\bibhang}}
    \def\newblock{\hskip .11em plus .33em minus .07em}
    \sloppy\clubpenalty4000\widowpenalty4000
    \sfcode`\.=1000\relax}}%
{\def\thebibliography#1{
  \list
  {\relax}{\setlength{\labelsep}{0em}
        \setlength{\itemindent}{-\bibhang}
        \setlength{\itemsep}{\parskip}          
        \setlength{\parsep}{0pt}
        \setlength{\leftmargin}{\bibhang}}
    \def\newblock{\hskip .11em plus .33em minus .07em}
    \sloppy\clubpenalty4000\widowpenalty4000
    \sfcode`\.=1000\relax}}

\newlength{\bibhang}
\let\@internalcite\cite
\def\cite{\@ifstar{\citey}{\citefull}}
\def\citefull{\def\astroncite##1##2{##1\ ##2}\@internalcite}
\def\citey{\def\astroncite##1##2{##1\ (##2)}\@internalcite}
\def\citeyear{\def\astroncite##1##2{##2}\@internalcite}
\def\citename{\def\astroncite##1##2{##1}\@internalcite}
\def\@citex[#1]#2{\if@filesw\immediate\write\@auxout{\string\citation{#2}}\fi
  \def\@citea{}\@cite{\@for\@citeb:=#2\do
    {\@citea\def\@citea{; }\@ifundefined       
       {b@\@citeb}{{\bf ??}\@warning              
       {Citation `\@citeb' on page \thepage \space undefined}}%
{\csname b@\@citeb\endcsname}}}{#1}}

\def\@cite#1#2{#1\if@tempswa #2\fi}       
\def\@biblabel#1{}

\def\astroncite#1#2{#1\ #2}
\makeatother

\begin{document}

\thesaurus{06(%
08.06.1; 
13.25.5; 
08.18.1; 
08.12.1; 
08.09.2)} 

\title{Rotational modulation of X-ray flares on late-type stars: T Tauri Stars and Algol}

\author{B. Stelzer\inst {1} \and R. Neuh\"auser\inst {1} \and
S. Casanova\inst {2} \and T. Montmerle\inst {2}}

\institute{Max-Planck-Institut f\"ur extraterrestrische Physik,
  Giessenbachstr.~1,
  D-85740 Garching,
  Germany \and
  Service d'Astrophysique, 
  CEA Saclay, F-91191 Gif-sur-Yvette, France}

\offprints{B. Stelzer, stelzer@xray.mpe.mpg.de}
\mail{B. Stelzer}
\titlerunning{Rotational modulation of X-ray flares}

\date{Received 23 October 1998 / Accepted 18 December 1998} 
\maketitle
 
\begin{abstract}
We present evidence for rotational modulation of X-ray flares by an
analysis of four outbursts on late-type stars. Two of the flares we
discuss are found in {\em ROSAT} observations of T Tauri Stars and were 
obtained between September and October 1991. A flare on the T Tauri
Star V773\,Tau was observed by {\em ASCA} in September 1995.
To this sample we add a {\em Ginga} observation 
of a flare on Algol observed in January 1989.

The structure of the X-ray lightcurves observed in this selection of 
flare events is untypical in that the maximum emission extends 
over several hours
producing a round hump in the lightcurve instead of a sharp peak.
We explain this deviation from the standard shape of a flare lightcurve 
as the result of a flare erupting on the back side of the
star and gradually moving into the line of sight due to the star's
rotation. Making use of the known rotational periods of the stars our model
allows to determine the decay timescale of the flares and the size of the
X-ray emitting volume according to the standard magnetic loop model. 
Spectral information, which is available in sufficient quality for the Algol
observation only, supports our proposition 
that changes of the visible volume are responsible for the observed time 
development of these flares.

\end{abstract}

\section{Introduction}\label{sect:intro}

After the first stellar X-ray flares were discovered less than 25 years ago on
dMe stars (\cite{Heise75.1}) it took almost another decade until the 
{\em Einstein observatory} ({\em EO}) detected similar events on young 
T Tauri Stars (TTS) (\cite{Montmerle83.1}). 
Nowadays, X-ray flares are known to be entertained on stars all 
over the H-R diagram (see \cite{Pettersen89.1} for a review).
The timescales and energetics
involved in flare events on different types of stars vary strongly
consistent with the observation 
that the level of X-ray activity decays with age. 

TTS are late-type pre-main sequence stars with typical age of 
$10^5 - 10^7$\,yrs
and rank among the most active young stars: energy outputs 
of up to $10^4$ times the maximum X-ray emission observed from solar flares 
have been reported from TTS outbursts.
Some of the largest X-ray flares ever observed
were discovered by {\em ROSAT} on the TTS LH$\alpha$~92 and P1724 
(\cite{Preibisch93.1}, \cite{Preibisch95.1}). The
energy released in these events ($> 10^{36}\,{\rm ergs}$) exceeds that
of typical TTS flares by two orders or magnitude. 
Before the detection of these giant events, the record of X-ray luminosity
was held for more than 10\,years by the TTS ROX-20, where 
$L_{\rm x} \sim 10^{32}\,{\rm ergs/s}$ were measured during 
an {\em EO} observation in February 1981 (\cite{Montmerle83.1}).
A superflare from
the optically invisible infrared Class~I protostar YLW\,15 
in $\rho$ Oph was
presented by \citey{Grosso97.1}, 
with the intrinsic X-ray luminosity
over the whole energy range being $10^{34}$ to $10^{36}$\,erg/s, depending
on the foreground absorption which is known to lie somewhere between 20 and
40 mag.

Although no model has been found yet that explains all aspects of
flaring activity,
the basic picture of all flare scenarios is -- in analogy to the sun --  
that of dynamo driven magnetic field loops that confine a hot, 
optically thin, X-ray emitting plasma (see \cite{Haisch91.1}
for a summary of flare phenomena). 
Quasi-static cooling of such coronal loops has been described by
\citey{Oord89.1}. 
An analysis of single flare events is of interest 
to determine physical parameters of the flaring region 
such as time scales, energies, temperature and plasma density, and
ultimately decide whether coronal X-ray emission of TTS is scaled-up 
solar activity, or whether interaction between the star and either a 
circumstellar disk or a close binary companion are partly responsible for
the X-ray emission.
 
In this paper we select a sample of four X-ray observations 
(three of TTS and one of Algol) which are 
in conflict with the standard modeling of the lightcurve as either a flare
characterised by a 
quick rise and subsequent exponential decay or as simple sine-like
rotational modulation of the quiescent emission. In the latter case X-ray
emission would be larger when the more X-ray luminous area is on the front
side of the star (directed towards the observer). Such kind of rotationally
modulated emission was observed in the TTS SR\,12 in $\rho$ Oph by
\citey{Damiani94.1}. 

We propose that the untypical shape of the
X-ray lightcurves we present is due to a flare event modulated by the 
rotation of the star. \citey{Skinner97.1} 
suggested rotational occultation of an X-ray flare
to explain the broad maximum and slow decay of a flare on V773\,Tau observed
by {\em ASCA}. While they model their data by fitting a sine function to the
lightcurve without allowing for an exponential decay phase 
(similar to \cite{Damiani94.1}), we start out
from a decaying flare and modify it by a time varying volume factor.
By this approach we 
take into consideration that the flare might be occulted by the star
during part of the observation and we are able to estimate 
the decay timescale of the lightcurve $\tau$ and the size of the emitting loop.
Such a model was first suggested by \citey{Casanova94.1},
and \citey{Montmerle97.1} 
classified the corresponding flare event as `anomalous'. A rotationally 
modulated flare was also
mentioned as possible interpretation of a flare-like event in P1724 
(\cite{Neuhaeuser98.1}), shown in our Fig.~\ref{fig:fit_P1724}, who advertised
the more detailed and quantitative treatment that we present in this paper. 

The outline of our presentation is as follows:
In Sect.~\ref{sect:data} we introduce the X-ray observations that we chose 
in view of the untypical broad maximum of their lightcurves.
A model that describes modulations of X-ray flares by the rotation of
the star is 
presented in Sect.~\ref{sect:model}. In Sect.~\ref{sect:applic} we 
explain the structure of the lightcurves from the observations introduced in 
Sect.~\ref{sect:data} by applying our model, and we 
summarize the results in Sect.~\ref{sect:conclusions}.

\section{The observations}\label{sect:data}

A summary of the observations analysed in this paper is given in
Table~\ref{tab:data}. The observations were obtained with different instruments 
onboard the X-ray satellites {\em ROSAT}, {\em Ginga}, and {\em ASCA}.
For information about the {\em ROSAT} instruments, the Position Sensitive
Proportional Counter (PSPC) and the High Resolution Imager (HRI), 
we refer to \citey{Truemper82.1}. 
The {\em Ginga} Large Area Counter (LAC) has been described by 
\citey{Turner89.1}, 
and a description of {\em ASCA} and its instrumentation can be found in 
\citey{Tanaka94.1}.
\begin{table}
\caption{Flare observations}\label{tab:data}
\begin{tabular}{lllll} \hline
Star & \hspace{-2mm} Instrument & \hspace{-1mm} Principal & \hspace{-2mm} date & \hspace{-2mm} time \\ 
     &            & Investigator & \hspace{-2mm} [UT] & \hspace{-3mm} [ksec] \\ \hline
Algol & \hspace{-2mm} {\em Ginga} LAC & \hspace{-1mm} Stern,\,R.\,A. & \hspace{-2mm} 12-14 Jan 89 & \hspace{-2mm} 37.9 \\
SR\,13 & \hspace{-2mm} {\em ROSAT} PSPC & \hspace{-1mm} Montmerle,\,T. & \hspace{-2mm} 07/08 Sep 91 & \hspace{-2mm} 19.9 \\
P1724 & \hspace{-2mm} {\em ROSAT} HRI & \hspace{-1mm} Caillault,\,J.\,P. & \hspace{-2mm} 02/03 Oct 91 & \hspace{-2mm} 28.1 \\
V773\,Tau & \hspace{-2mm} {\em ASCA} SIS0 & \hspace{-1mm}
Skinner,\,S.\,L. & \hspace{-2mm} 16/17 Sep 95 & \hspace{-2mm} 19.8 \\ \hline
\end{tabular}
\end{table}

The classical TTS (CTTS) SR\,13 was detected in X-rays by 
\citey{Montmerle83.1}. 
It is located in the $\rho\,{\rm Oph}$ cloud at 
position  $\alpha_{2000}=16^{\rm h} 28^{\rm m} 45^{\rm s}.3$ , 
$\delta_{2000}=-24^\circ 28' 17.0''$.  
On a speckle imaging survey SR\,13 was discovered to be a binary system 
(\cite{Ghez93.1}) with $0.4''$ separation, which remains unresolved 
in the PSPC observation of 1991 September 07/08 we present here.
The period of SR\,13 is unknown. However, as shown below, a period of
 $\sim 3 - 6$ days is consistent with the rotating X-ray flare model.

P1724 is a weak line TTS (WTTS) located 15 arc min north of the Trapezium
cluster in Orion ($\alpha_{2000}=5^{\rm h}35^{\rm m}4^{\rm s}.21$, 
$\delta_{2000}=-5^\circ8'13''.2$).
\citey{Neuhaeuser98.1} 
confirm the rotational period of 5.7\,d first discovered by 
\citey{Cutispoto96.1} 
applying two independent numerical period search methods 
on the V-band lightcurve. In addition, they report systematic variations of the
X-ray count rate of P1724 during an observation with the {\em ROSAT} HRI in
October 1991. However, they cannot find any rotational modulation in the
X-ray data. \citey{Neuhaeuser98.1} 
also find no indications for a
circumstellar disk nor a close binary companion.

The WTTS V773\,Tau is a double-lined 
spectroscopic binary (\cite{Welty95.2}) which is located in 
the Barnard\,209 dark cloud at optical position 
($\alpha_{1950}=4^{\rm h}11^{\rm m}07.29^{\rm s}$, 
$\delta_{1950}=28^\circ04'41.2''$). Upper limits for the rotation period
derived by \citey{Welty95.2} 
are $2.96\,{\rm d}$ and $2.89\,{\rm d}$ for
the K2 and K5 components respectively. These estimates are somewhat lower than 
the values previously reported by \citey{Rydgren83.1}.
In the {\em ASCA} observation obtained on 1995 September 16/17 
the V773\,Tau binary system is not resolved from
the classical TTS (CTTS) FM\,Tau, which lies at an offset of $\sim 38''$. 
For a more detailed discussion of this {\em ASCA} observation we refer 
to \citey{Skinner97.1}.

Algol is a triple system, with an inner close binary of period 2.87\,d
comprising a B8\,V primary (Algol A) and an evolved K2\,IV secondary (Algol B).
In January 1989 {\em Ginga} observed a large flare from the Algol
system (presumably Algol B; \cite{Stern92.1}). The shape of the Algol 
lightcurve resembles that of the TTS flares discussed before. 
Therefore, we decided to include this observation in our sample, 
although the Algol system represents a different class of flare stars.

We use the Extended Scientific Software Analysis System 
(EXSAS, \cite{Zimmermann95.1}) to analyse
the two {\em ROSAT} observations, which were obtained from the 
{\em ROSAT} Public Data Archive. To take account of 
possible time variations in the background,
the background count rate was computed for each satellite orbit and
subtracted accordingly from the measured lightcurve. 
{\em Ginga} data were kindly supplied to us in computer readable format by 
Bob Stern, while Steve Skinner 
provided us the {\em ASCA} lightcurve of V773\,Tau.

\section{Rotational effects on flare lightcurves and spectra}\label{sect:model}

The X-ray lightcurve during a flare event is commonly described by a steep,
linear rise followed by an exponential decay. The e-folding time $\tau$
of the decay varies greatly between several minutes to hours depending on
the nature of the flaring star. Disagreement
prevails on the question whether the apparent quiescent emission might
be attributed to continuous, unresolved short timescale activity.
The detection of a high temperature spectral component in the 
quiescent spectrum might hint at such low-level flaring (\cite{Skinner97.1}). 

\begin{figure}
\begin{center}
  \resizebox{5cm}{!}{\includegraphics{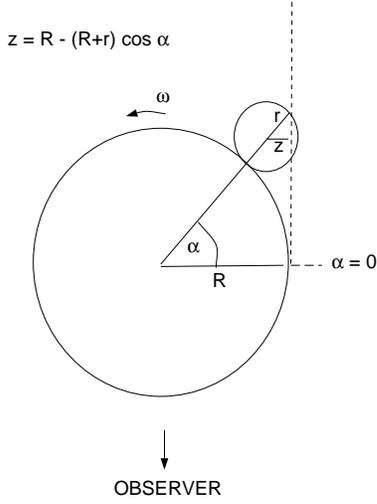}}
\caption{Top-view of the rotational plane of a flaring star. $R$ is the
radius of the star which rotates counterclockwise with angular 
velocity $\omega$. 
The flaring volume is represented by the smaller circle of radius $r$. 
The angle $\alpha (t)$ is defined to be zero at the time 
when the plasma loop starts to disappear from the observer's view as a
consequence of the rotation of the star.}
\label{fig:model}
\end{center}
\end{figure}

Several flares have been observed that do not match the typical 
appearance: instead of displaying a sharp peak, the lightcurves of 
this type of events are characterized by smooth variations 
around maximum 
emission that sometimes goes along with a slower rise as compared to 
standard flare events. 
In these cases the shape of the lightcurve can be reproduced by 
taking account of the
rotation of the star. Flares that erupt on the backside of the star become
visible only gradually as the star rotates and drags the plasma loop
around. The visible flare volume is thus a function of time which modulates
the exponential decay. The scenario we have in mind, and that we will 
refer to as the `rotating flare model' henceforth, 
is sketched in Fig.~\ref{fig:model}.

\begin{figure}
  \resizebox{\hsize}{!}{\includegraphics{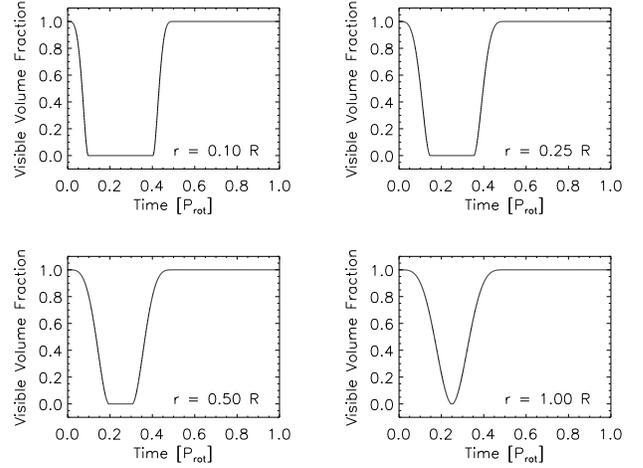}}
\caption{Visible fraction of the plasma loop volume approximated by our
rotating-flare model of Fig.~\ref{fig:model} as a function of time for 
different values of the radius fraction $f=\frac{r}{R}$.}
\label{fig:vol}
\end{figure}

For simplicity the emitting plasma loop is approximated by a sphere 
anchored on the star's surface. The fraction of the loop volume 
which is visible to the observer is given by
\begin{equation}
V (r,t) =  \frac{1}{\frac{4}{3} \pi r^3} \int^{r}_{R - (R+r) \cdot \cos{\alpha (t)}}{\pi (r^2-x^2)\,{\rm d}x}
\label{eq:vol}
\end{equation}
where $R$ is the radius of the star and $r$ 
the radius of the spherical plasma
loop. The time dependency of $V$ is hidden in $\alpha$, the 
angle between the current position of the flaring volume and the position 
where a flare just begins to become occulted by the star. 
Note that Eq.~(\ref{eq:vol}) does not hold for 
rotational phases during which the flaring volume is either completely
behind ($\alpha \sim \frac{\pi}{2}$) or completely in front of the star 
($\alpha \in [\pi,2 \pi]$). The time dependency of $\alpha$ in 
Eq.~(\ref{eq:vol}) depends on whether the loop is disappearing or reappearing
and is given by
\begin{equation}
\alpha (t) = \left\{ \begin{array}{cc}
			\frac{2 \pi t}{P_{\rm rot}} & \mbox{for $0 \leq \alpha (t) < \phi_{\rm crit}(f)$} \\
			\pi - \frac{2 \pi t}{P_{\rm rot}} & \mbox{for $\pi
			- \phi_{\rm crit}(f) < \alpha (t) \leq \pi$}
	              \end{array}
	      \right.
\label{eq:alpha}
\end{equation}
where $\phi_{\rm crit}$ is the critical phase at which the plasma volume
has just disappeared. Eq.~(\ref{eq:vol}) is not valid any more until the
loop reaches phase $\pi-\phi_{\rm crit}$ and begins to move into the line 
of sight again. $\phi_{\rm crit}$ is a function of the relative size
of the radius of the flaring volume and the radius of the star, 
$f=\frac{r}{R}$.
The visible fraction of the plasma volume as a function of time is plotted
for different values of the radius ratio $f$ in Fig.~\ref{fig:vol}.

Our model is based on several simplifying assumptions concerning the flare
geometry. First, we imply that we look directly onto the rotational plane,
i. e. $i \sim 90^\circ$, and that the flare takes place at low latitudes.
Flares that erupt in polar regions, in contrast, in the configuration 
of Fig.~\ref{fig:model} would remain partially visible during the whole 
rotation period. Furtheron, Eq.(\ref{eq:vol}) does not take account of the
curvature of the star. We content ourselves with these approximations
because, given the present quality of the data, further
sophistication of the model seems to be unnecessary. 

Making use of the configuration described above, 
for a flare which is observed while the flaring region turns up from the
backside of the star, the X-ray lightcurve
can be modeled by
\begin{equation}
I_{\rm cps} = I_{\rm q} + I_0 \cdot \exp(-t/\tau) \cdot V (r,t)
\label{eq:fit}
\end{equation}
where $I_{\rm q}$ is the quiescent X-ray count rate of the star, 
$I_0$ the strength of the outburst, $\tau$ the decay
timescale of the count rate and $V(r,t)$ the visible fraction of the 
volume of the plasma loop given by Eq.~(\ref{eq:vol}) for values of
$\alpha$ within the allowed range, and by 0 or 1
for angles $\alpha$ outside the intervals of Eq.~(\ref{eq:alpha}).

The hump-like shape of the lightcurves we will discuss in the next section can
be reproduced if the visible volume $V$ {\em increases} during the
first {\em observed} part of the flare, ie. 
$\alpha \geq \pi - \phi_{\rm crit}$. Three critical moments determine the 
rotating flare event: the
time of outburst, the time when the flare region passes phase 
$\pi - \phi_{\rm crit}$ and begins to move into the line of sight, and the
time at which the observation started. In the next paragraphs the relation
between these times will be examined.

\begin{figure}
  \resizebox{\hsize}{!}{\includegraphics{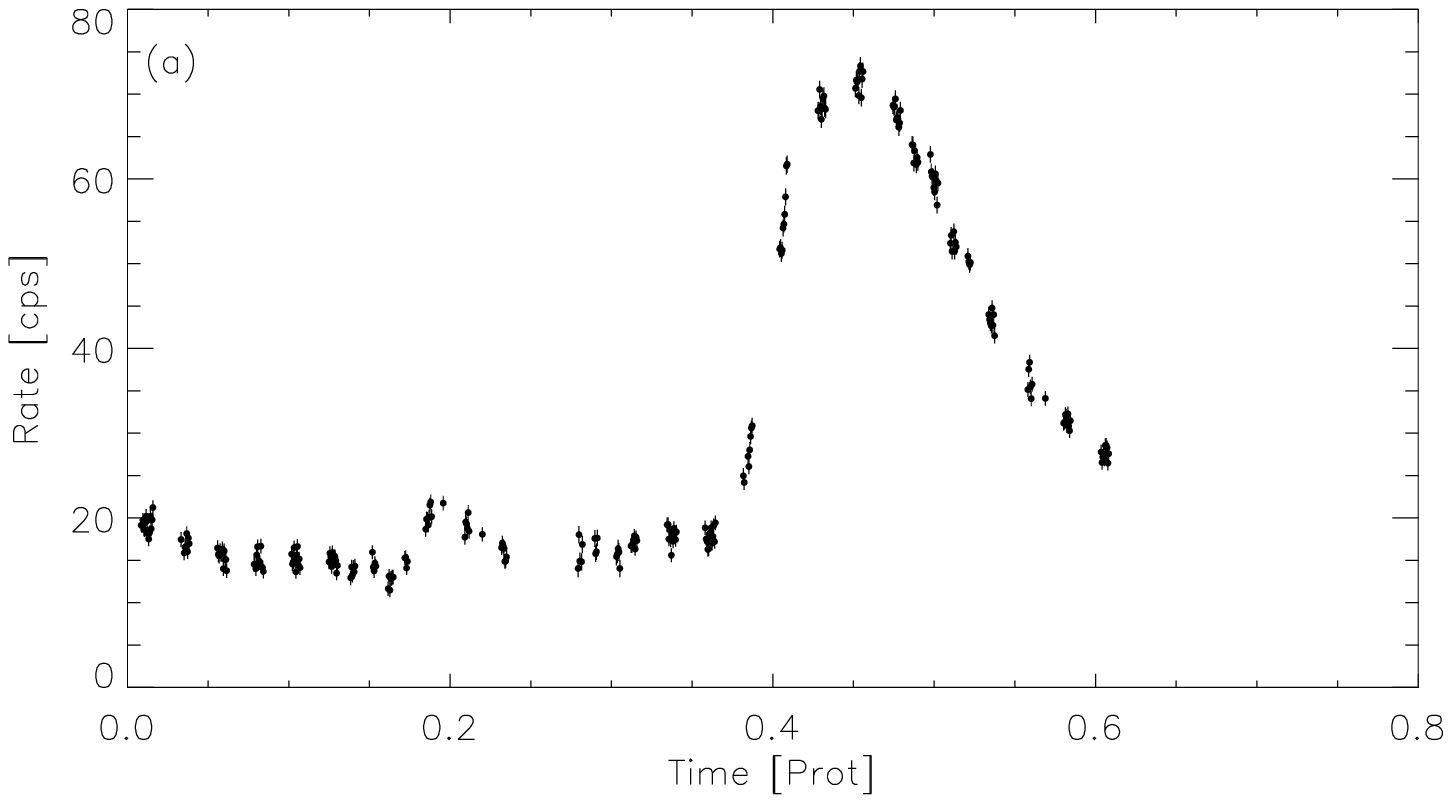}}
  \resizebox{\hsize}{!}{\includegraphics{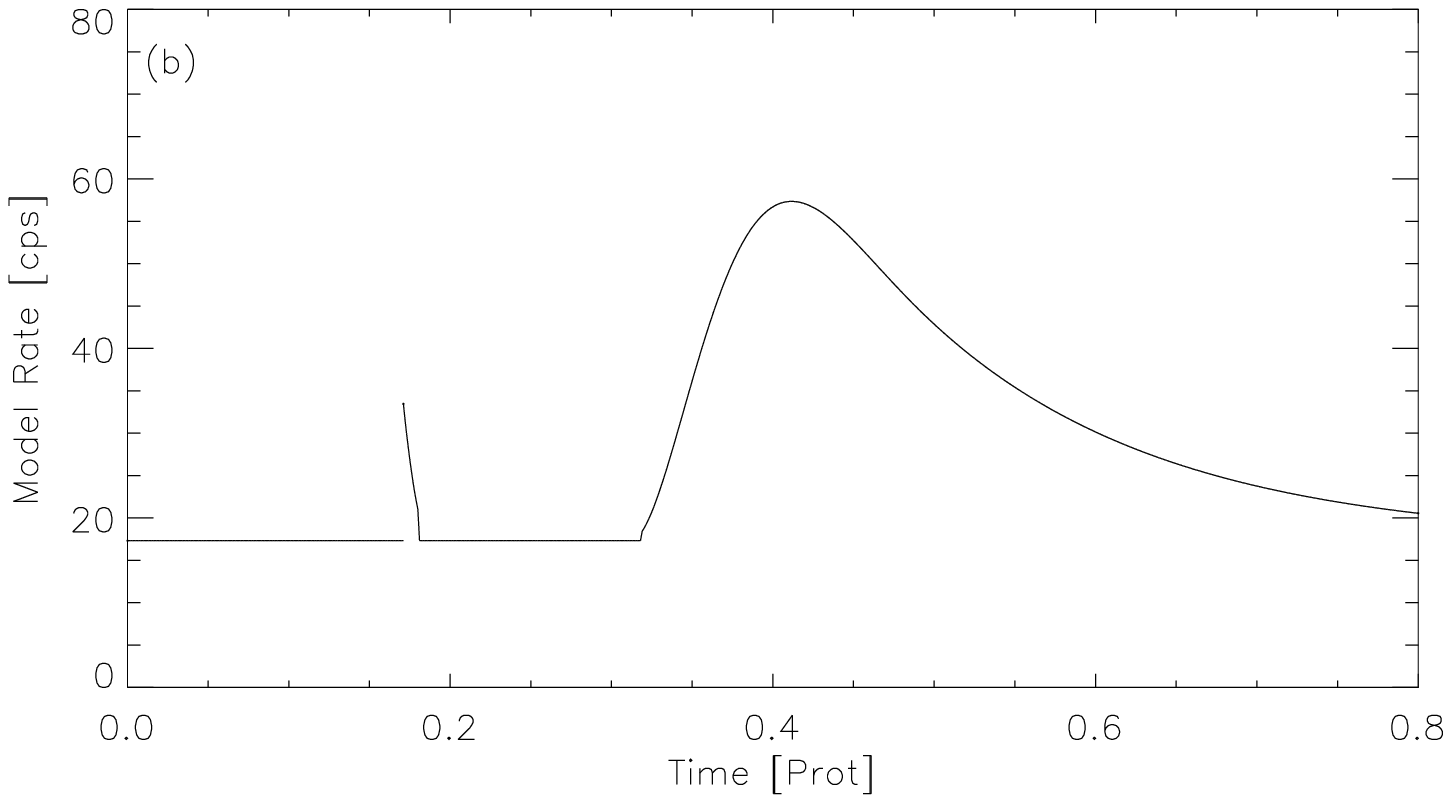}}
  \resizebox{\hsize}{!}{\includegraphics{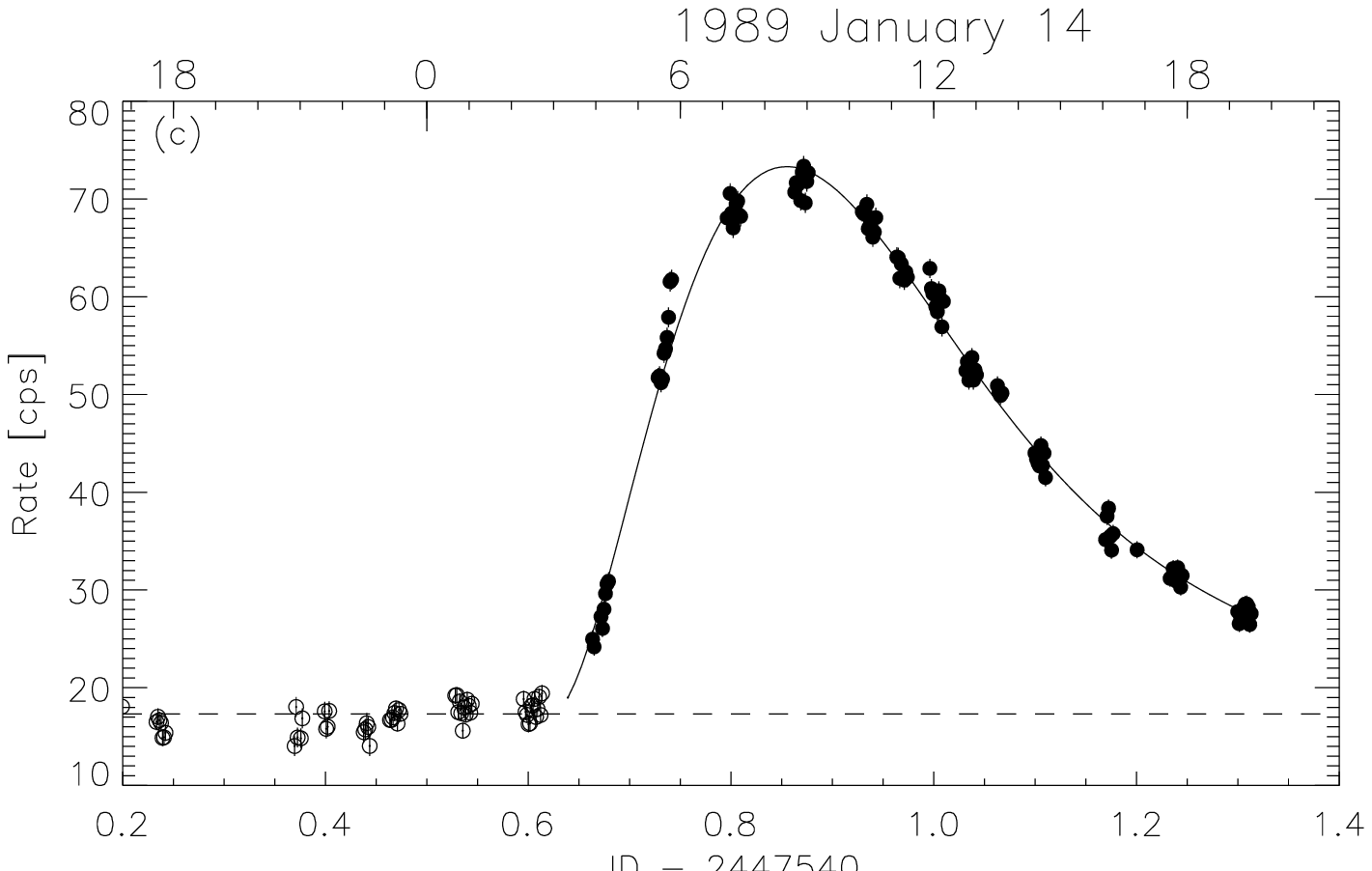}}
  \caption{{\em GINGA} lightcurve of the Algol flare in comparison with the
  rotating-flare model: (a) Algol lightcurve including the short event 
  13\,hours before the large flare, (b) theoretical shape of 
  a lightcurve representing a strong flare which is occulted 
  immediately after its outburst and reappears later,
  and (c) large Algol flare overlaid by our best fit model
  curve. Binsize: 128\,s (1\,$\sigma$ uncertainties)}
  \label{fig:theolcs}
\end{figure}

First, an offset between flare outburst and the time when it becomes visible
to an observer (at $\pi - \phi_{\rm crit}$) 
might be present, when the flare takes place on the
occulted side of the star.
In our model such a time offset $\Delta t$ 
contributes only to the normalization of the exponential
$I_0 = I_{\rm intr} \cdot \exp(-\Delta t/\tau)$ and cannot be separated 
from the intrinsic brightness $I_{\rm intr}$ of the outburst.
The upper limit for $\Delta t$ is given by
\begin{equation}
\Delta t_{\rm max} = (0.5 - 2 \phi_{\rm crit}) P_{\rm rot}
\end{equation}
since for larger time offsets the flare would have been observed 
also at $\alpha < \phi_{\rm crit}$, that is before its occultation. 
Given the rotational periods of several days, 
$\Delta t_{\rm max}$ exceeds the typical decay timescale for TTS flares
(of a few hours).
However, 
from an observational point of view it is impossible to exclude that the
flares occurred already before they rotated away, because
data extending over several hours before the reappearance of the flare
are not available for the lightcurves analysed here, except in the case of 
Algol. Indeed, at first glance the
combination of the two phases of enhanced count rate in the Algol observation 
(see Fig.~\ref{fig:theolcs}\,a) 
looks similar to what is expected 
to be seen from {\it one} very long flare that
disappeared behind the star shortly after outburst and reappeared
half a rotational cycle later still displaying a strong count rate
enhancement. Figure~\ref{fig:theolcs}\,b gives an example of a theoretical
lightcurve for a flare which is occulted right after its outburst
and whose duration is more than half the rotation period.
However, our attempt to model
the complete Algol lightcurve from Fig.~\ref{fig:theolcs}\,a 
by such a single temporary occulted flare was not successful
because the model restrictions concerning the relative strength of the pre-
and post-occultation part of the flare are not met by the Algol lightcurve.
Thus we can rule out offsets larger than $\Delta t_{\rm max}$ for the
Algol observation discussed in this paper, 
and the short rise in count rate observed before the large flare must be
due to an independent event.

Second, the start of the observation of a flare event, which is characterised by
a beginning enhancement of the observed count rate, can differ from the
time at which the outer edge of the plasma loop emerges from the back of
the star due to gaps in the data stream. So, strictly speaking, 
another offset $\delta t$
has to be included when fitting the `rotating flare model' to the data to
take account of a possible delay of the observation with respect to flare
phase $\pi - \phi_{\rm crit}$. 
Such an additional parameter that allows to determine the
rotational phase of the flare region at the beginning of the observed rise 
is needed to obtain acceptable fits for the flares on Algol and V773\,Tau.
For the {\em ROSAT} observations (of SR\,13 and P1724), however, an offset
$\delta t$
does not improve the fit due to the low statistics of the data.
We note here, that the observed flare rise is only {\em apparent}
according to our model: The star is assumed to have flared
(and thus exhibited its maximum emission) well before the {\em observed}
maximum, and the count rate is low at that time only due to the fact that
the flaring volume has not yet become visible.

The enhanced X-ray emission during flare events is produced by a hot plasma
which has been heated to temperatures of $10^6\,{\rm K}$ and
above. Optically thin plasma models show, 
when applied to spectra representing different stages of the flare, 
that after the outburst the temperature drops exponentially 
to the quiescent level. The  
temperature observed for a rotationally modulated flare should thus be highest
during the phase where the flare emerges from the backside of the star 
when the observed lightcurve has not yet reached its maximum. The emission
measure, on the other hand, being a volume related parameter is expected to
show a time evolution similar to the lightcurve.

\section{Application of the model}\label{sect:applic}

We fit the model of Eq.~(\ref{eq:fit}) to that part of the lightcurves
from the observations introduced in Sect.~\ref{sect:data} that are by visual
inspection identified with the outburst due to their enhanced count rates. 
Except for V773\,Tau (see Fig.~\ref{fig:fit_V773Tau}) 
none of these lightcurves 
(Fig.~\ref{fig:theolcs}\,(c), Fig.~\ref{fig:fit_SR13}, and 
Fig.~\ref{fig:fit_P1724}) can be explained by
simple sine-like variations due to rotational modulation of the 
quiescent emission. There is always an additional feature present, 
namely a flare.
The lightcurves discussed here are characterised by a concave shape of the 
(e-folding) decay phase typical for flares (whether or not rotationally
modulated),
while a simple sine-like rotational modulation of quiescent emission always
produces a convex shape in the decay part.

In all cases we examined, the quiescent count rate is held fixed on its 
average pre-flare value. Thus, three free parameters have to be adjusted
to the data: the strength of the flare, $I_0$ in cts/s, the decay timescale
of the lightcurve, $\tau$, and the radius $r$ of the flaring volume relative to
the stellar radius, $R$. An additional freedom allowing for 
an offset $\delta t$ between rotational phase 
$\pi - \phi_{\rm crit}$ of the flare 
and the {\em apparent} outburst of the flare, which is observed as a rise
in count rate, is used for the modeling of the {\em Ginga} (Algol) and 
{\em ASCA} (V773\,Tau) lightcurves
(see the explanation in the previous section).

The rotational periods $P_{\rm rot}$ of our sample stars 
were known from optical photometry, except for the case of SR~13. In
principle, $P_{\rm rot}$ could be included as a further free parameter in
the fit. But with this additional freedom the fit does not result in a unique
solution, as we will show in the example of SR\,13, and thus $P_{\rm rot}$
may not be uniquely determined from our model. For Algol we assume
synchronous rotation. 

Our best fit results will be discussed in detail in the following subsections.
The best fit parameters for all flares are listed in
Table~\ref{tab:bfparams} together with
the rotation periods and measured quiescent count rates.
\begin{table*}
\caption{Best fit parameters of the rotating-flare model and 1\,$\sigma$ 
     uncertainties. No uncertainties are given for the {\em ROSAT}
     observations (see text). Dots indicate that no uncertainties 
     could be determined within the limits of the parameters given by 
     model restrictions (ie. $0 < r < 1$).The rotational
     periods were fixed on the values given in column\,2 except for SR\,13
     whose rotational period is unknown: $^a$ \protect\cite{Hill71.1} 
     (assuming synchronous rotation of Algol B), 
     $^b$ \protect\cite{Cutispoto96.1}, and $^c$ \protect\cite{Welty95.2}. The 
     quiescent count rate, $I_{\rm q}$, was determined from the pre-flare
     data except for V773\,Tau.}\label{tab:bfparams}
\begin{tabular}{lccccccc} \hline
Star & $P_{\rm rot}$ & $I_{\rm q}$ & $I_0$ & $\tau$ & $r$ & ${\delta t}$ & $\chi_{\rm red}^2$\,(dof) \\
     & [d]           & [cps]           & [cps] & [h]    & [$R_*$] & [h] & \\ \hline
Algol & $2.87^a$ & $17.32 \pm 0.02$ & $207.8^{+72.2}_{-64.1}$ &
     $~5.2^{+0.1}_{-0.4}$ & $0.55^{+...}_{-0.12}$ & $0.6^{+0.8}_{-0.1}$ & $3.05\,(107)$ \\ 
SR\,13 & $3$ \hspace{0.5cm} & $0.03$ \hspace{0.8cm} & ~$0.58$ \hspace{0.3cm} & $4.1$ \hspace{0.4cm} & $0.65$ \hspace{0.6cm} & & $1.23\,~(36)$ \\
       & $6$ \hspace{0.5cm} & $0.03$ \hspace{0.8cm} & ~$2.58$ \hspace{0.3cm} & $2.9$ \hspace{0.4cm} & $0.65$ \hspace{0.6cm} & & $1.16\,~(36)$ \\
P1724 & $5.7^b$ \hspace{0.3cm} & $0.04$ \hspace{0.8cm} & ~$0.25$ \hspace{0.3cm} & $4.8$ \hspace{0.4cm} & $0.10$ \hspace{0.6cm} & & $0.99\,~(36)$ \\
V773\,Tau$^*$ & $2.97^c$ & $0.10 \pm 0.02$ & $~~2.1^{+1.1}_{-0.6}$ & $21.9^{+7.8}_{-8.3}$ & $0.22^{+...}_{-...}~$ & $3.9^{+...}_{-2.1}$ & $1.45\,(149)$ \\ \hline
\multicolumn{8}{l}{$^*$ Fit includes an additional X-ray emitting region (see
     text)} \\
\end{tabular}
\end{table*}
Note, however, that 
the model depends to some degree on the initial parameters,
and the parameters are not well determined due to correlations, such that similar solutions are obtained for different combinations of parameter values.
For the flares on Algol and V773\,Tau we computed 90\,\% confidence levels
for the best fit parameters according to the method described by 
\citey{Lampton76.1}. The low statistics in the data of the lightcurves
of SR\,13 and P\,1724 do not allow to apply this method. We, therefore, 
do not give uncertainties for the best fit parameters of these events.

\subsection{Algol}\label{subsect:Algol}

A two day long continuous {\em Ginga} observation of Algol in January 1989
 (first presented by \cite{Stern90.1}, 1992) 
includes a large flare event.
Secondary eclipse begins during the decay of that flare, 
but it seems to affect the count rate only marginally.
Preceding the large flare, primary eclipse and a small flare are observed
(see discussion in Sect.~\ref{sect:model}). 
We therefore base our estimate for the quiescent
emission on the time between the two flare events, i.e. immediately before
the rise phase of the large outburst which marks the onset 
of the time interval to which we apply the
`rotating-flare model'. 
From fitting Eq.~(\ref{eq:fit}) to the data after ${\rm JD}\,2447540.65$ in 
Fig.~\ref{fig:theolcs}\,(c) we obtain a best fit $\chi^2_{\rm red}$ of 5.34 for
108 degrees of freedom. The fit can be significantly improved when the
critical phase $\pi - \phi_{\rm crit}$ is allowed to vary around the start
of observation as explained in Sect.~3 
($\chi^2_{\rm red}=3.05$ for $107$ dofs), and all stages of 
the flare are well represented by the model. Although this value of 
$\chi^2_{\rm red}$ is still far from representing an excellent fit,
the ability of the model to reproduce the overall shape of 
the X-ray lightcurve is convincing.

\begin{figure}
  \resizebox{\hsize}{!}{\includegraphics{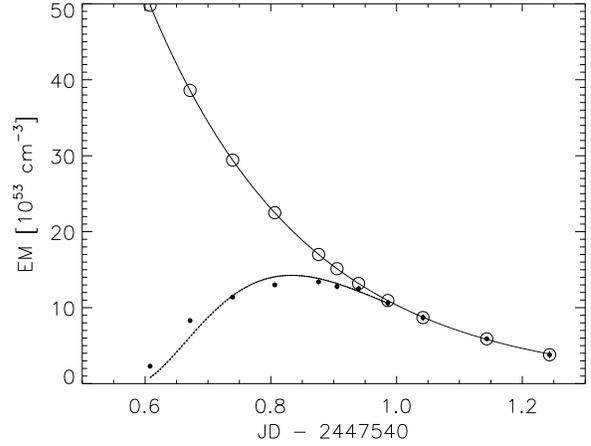}}
  \caption{Emission measure during the large flare on Algol. Small, filled
  circles represent the spectral results of \protect\citey{Stern92.1}. 
Large, open circles are our extrapolation of the exponential decay fitted
  to the last three data points of \protect\citey{Stern92.1}. 
The lower curve denotes the extrapolated emission measure after correction for the time dependence of the volume.}
  \label{fig:em}
\end{figure}

A detailed spectral analysis of the flare event on Algol was undertaken by
\citey{Stern92.1}. The emission measure $EM$ they obtained from a thermal
bremsstrahlung spectrum + Fe line emission for 11 time-sliced
spectra covering all phases of the flare is displayed in
Fig.~\ref{fig:em}. According to the best fit of our `rotating flare model'
to the lightcurve, the flare volume has become almost completely visible
($V>0.98$) around ${\rm JD}\,2447541$, i.e. about 10 hours after the rise
in count rate was observed to set in.
The exponential decay of the emission measure 
for the last three values of Fig.~\ref{fig:em} can thus be 
extrapolated to the previous part of the observation 
to find the values for the emission measure 
intrinsic to this flare event. The observed emission measure during 
the rotationally
dominated beginning of the flare is then fairly well reproduced by correcting
the extrapolated values for the time dependence of the volume 
($EM \sim n_{\rm e}^2 V$), where we neglect possible 
variations of the plasma density $n_{\rm e}$.
Thus, in contrast to the observation, the actual emission measure 
of the flare event
is highest at the onset of the flare at ${\rm JD} \sim 2447540.6$ 
and it decays simultaneously with
the count rate ($\tau_{\rm EM} = 6.45 \pm 0.67\,{\rm h}$) 
due to a decrease of $n_{\rm e}$ or shrinking loop volume.
The good agreement between the emission measure observed by 
\citey{Stern92.1} 
and the values expected from our model (see Fig.~\ref{fig:em}) 
provide convincing evidence that the 
application of the `rotating-flare model' is justified for this flare.
The development of the temperature during the flare (see \cite{Stern92.1}) 
does not show the characteristic hump shape, but is close to a pure
exponential decay as expected for volume unrelated parameters.

The values of the flare parameters ($\tau$, $r$) resulting from the fit 
of our model to the lightcurve are similar to those derived from normal
(ie. neither occulted nor rotationally modulated) 
Algol flares observed by various instruments. \citey{Ottmann96.1}
summarize the characteristic parameters of three Algol flares
(see their Table~5): the decay timescale seems to vary by one order of
magnitude between $\sim 3$ and $\sim 36$ hours, while the loop length
found from standard loop modeling extends from $\sim 0.5 - 2$ stellar 
radii. We conclude that modelling the January 1989 X-ray flare on Algol in
terms of rotational modulation yields flare properties which are perfectly
consistent with those of other X-ray flares.

\subsection{SR\,13}\label{subsect:SR13}

\citey{Casanova94.1} discusses the similarity between a flare of SR\,13 
observed by the {\em ROSAT} PSPC and
the Algol flare analysed in the previous subsection. Besides the absolute
values of the count rate which is by a factor of 500 higher for Algol
(note, that the observations were performed by different instruments and,
therefore, the differences in count rate are no direct measure for the
differences in flux),
the shape of the SR\,13 flare is very similar to that of the flare on Algol.

\begin{figure*}
  \parbox{15cm}{
  \parbox{7.5cm}{
  \resizebox{\hsize}{!}{\includegraphics{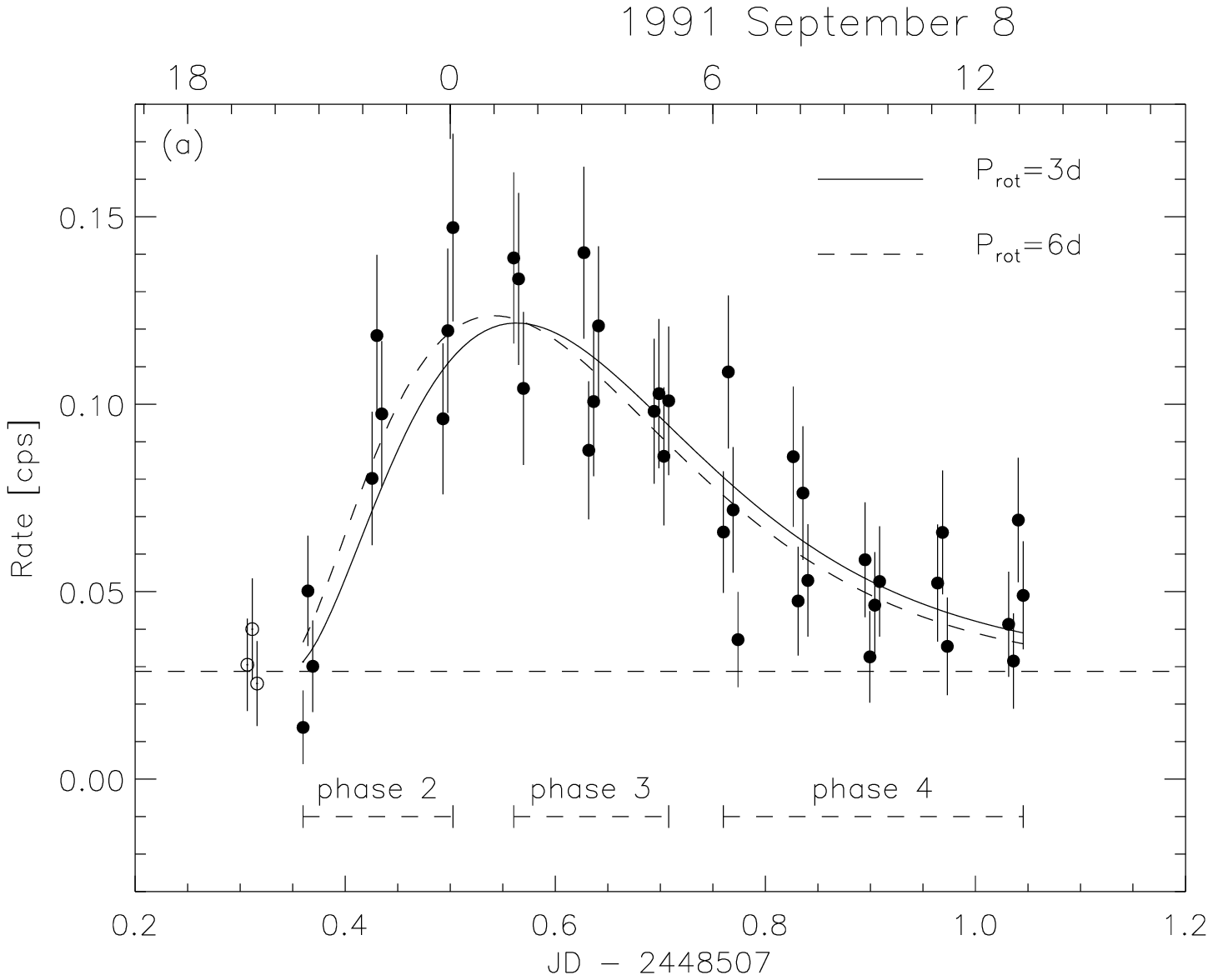}}
  \resizebox{\hsize}{!}{\includegraphics{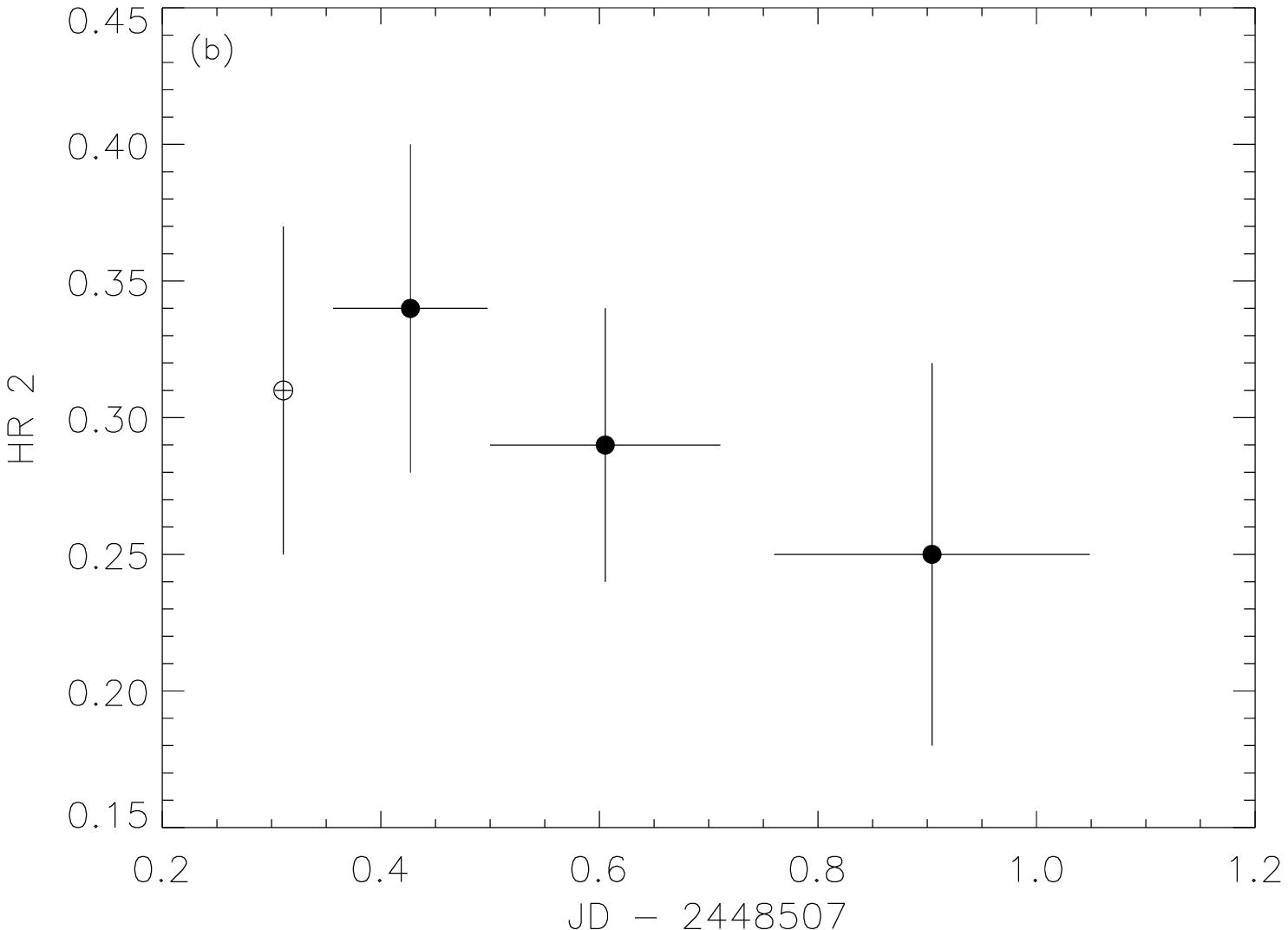}}}
  \parbox{7.5cm}{
  \resizebox{\hsize}{!}{\includegraphics{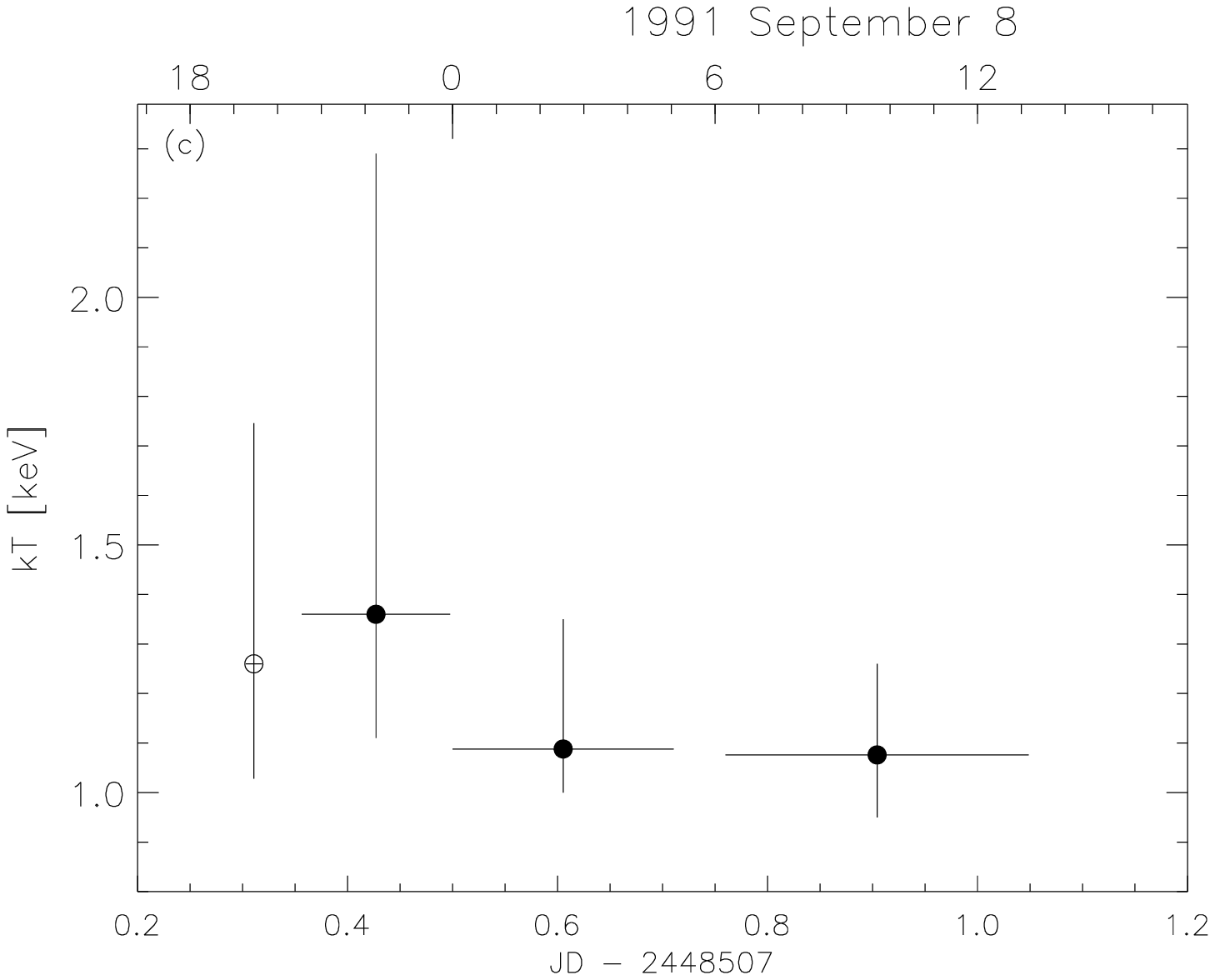}}
  \resizebox{\hsize}{!}{\includegraphics{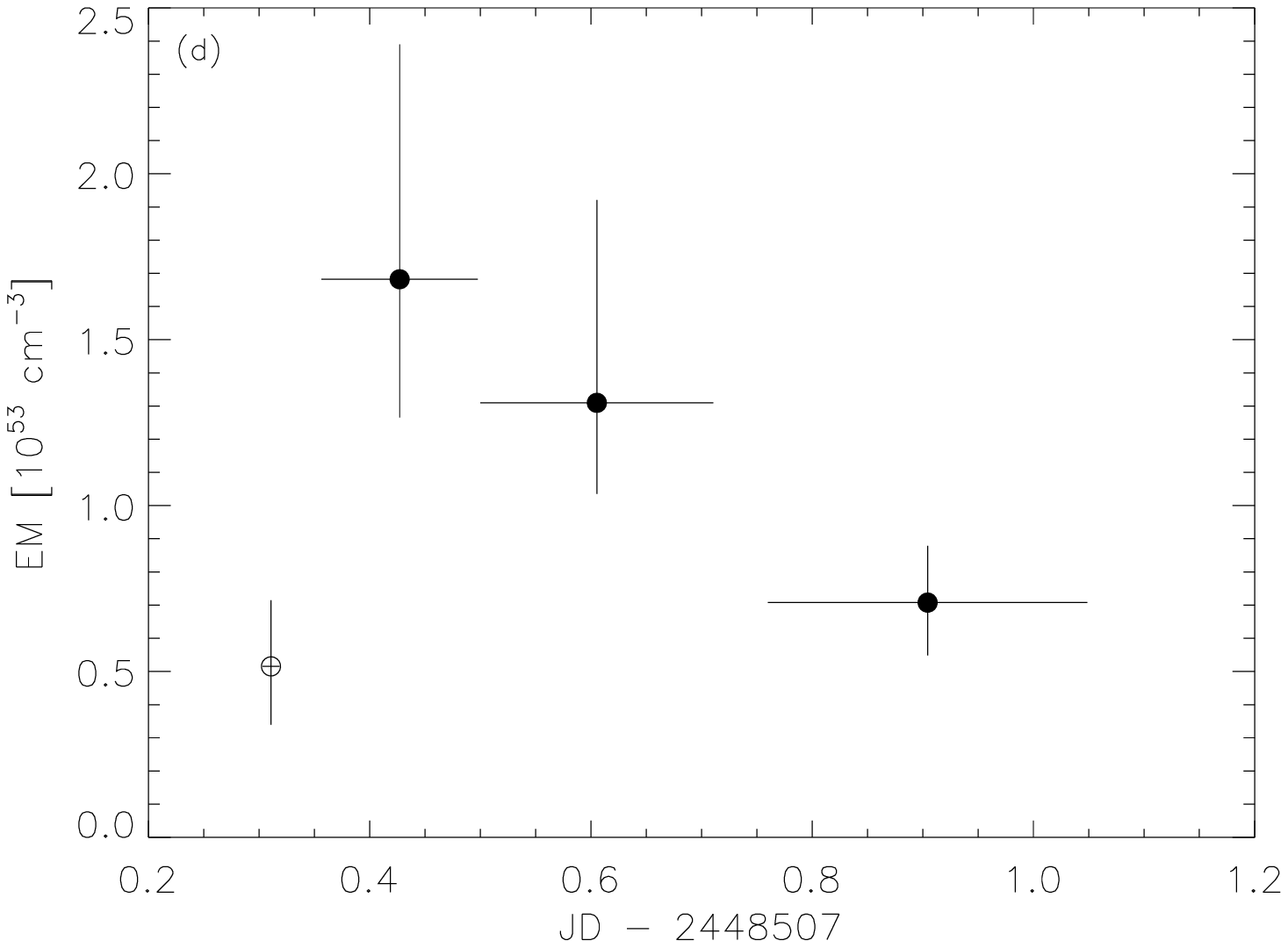}}}
  }
  \caption{{\em ROSAT} PSPC observation of a flare on SR\,13: (a) X-ray
  lightcurve (400\,s bins, 1\,$\sigma$ uncertainties)
  and best fit model. Data used in the fitting process 
  are represented by filled circles. Open circles are pre-flare data.
  (b) Time evolution of the hardness ratio {\em HR2} during the flare on
  SR\,13. Note, that the first value (open circle) was determined from the data
  of a 1991 March observation.
  Time dependency of the temperature (c) and emission measure (d) of a
  Raymond-Smith model spectrum display large uncertainties (shown are
  90\,\% confidence levels).}
  \label{fig:fit_SR13}
\end{figure*}

The rotational period of the CTTS SR\,13 is unknown to the present.
We determined the quiescent emission of SR\,13 
from the pre-flare data of the first satellite orbit. Our 
attempt to find the rotational period from the modeling of the flare 
according to Eq.~(\ref{eq:fit}) with $P_{\rm rot}$ a free parameter 
failed, since the uncertainties in the data 
do not allow to distinguish between different fit solutions.
In Fig.~\ref{fig:fit_SR13} (a)
we overlay the data points by two solutions of the
`rotating-flare model', one was found assuming a period of 3\,d, the other
one corresponds to twice that period. 

A detailed spectral analysis of this specific flare event similar to the one 
carried out for the Algol flare (see \cite{Stern92.1} and 
Sect.~\ref{subsect:Algol}) is not practicable due to the low number of
counts. To underline the difficulty in evaluating the spectral information
for the flare 
on SR\,13, we briefly discuss the results from our attempts to fit a
Raymond-Smith model (\cite{Raymond77.1}) 
to the spectra during four stages of the flare that were 
defined in the following way: phase 1 is given by the quiescent stage,
phase 2 is the observed, {\em apparent} flare rise, and phases 3 and 4 
correspond to the observed decay. The three flare time intervals are marked in 
Fig.~\ref{fig:fit_SR13}\,(a). The quiescent spectrum was computed
from an earlier observation obtained in 1991 March 05-10 by the {\em ROSAT}
PSPC due to the scarcity of 
non-flare data in the September observation. 

A two-temperature Raymond-Smith model was needed to obtain
acceptable fits with $\chi^2_{\rm red}<1.4$ for each of the four
phases, where we held the temperature
of the softer component fixed at $kT=0.25\,{\rm keV}$. The graphs in 
Fig.~\ref{fig:fit_SR13}\,(c) and (d) display the best fit values for the
temperature and emission measure of the hotter component. 
The large uncertainties of the best fit values shown in Table~\ref{tab:bfparams}
prohibit a spectral study with better time resolution, but having only
four time bins to define the spectral evolution, 
the decay of the emission measure
after the flaring volume became visible could not be pinned down, 
and thus a check of the `rotating
flare model' by modeling of the time development of the emission measure
is not possible for this flare on SR\,13. 
A slight indication for cooling is present in the evolution of 
$kT$ during the flare suggesting that the actual outburst might in fact
have occurred as early as during the second phase.

In cases of insufficient data quality hardness ratios may be used to 
give a clue to spectral properties. \citey{Neuhaeuser95.1} 
showed that the {\em ROSAT} hardness ratio {\em HR2} 
(see \cite{Neuhaeuser95.1} for a 
definition) is related to the temperature
of the plasma (see their Fig.~4). We computed {\em HR2} for the four different
time intervals defined above. 
The time evolution of the hardness ratio {\em HR2} is
displayed in Fig.~\ref{fig:fit_SR13} (b). The decreasing {\em HR2} during
the last three intervals supports the decline in temperature measured in
the spectra and presents further evidence for cooling.

To conclude, the results on the SR\,13 lightcurve, while having an admittedly
reduced statistical significance, are 
fully consistent with an interpretation
in terms of flare cooling combined with rotational modulation.

\subsection{P1724}\label{subsect:P1724}

\begin{figure}
  \resizebox{\hsize}{!}{\includegraphics{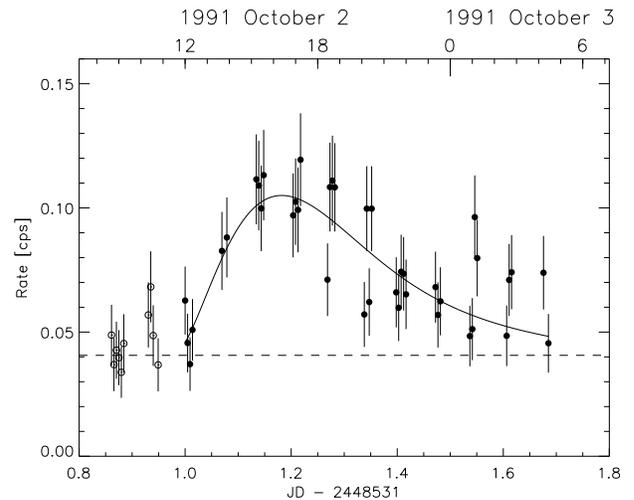}}
  \caption{{\em ROSAT} HRI lightcurve of the flare on P1724 (400\,s bins, 1\,$\sigma$
  uncertainties) and best fit model. The meaning of the plotting
  symbols is the same as in Fig.~\ref{fig:fit_SR13}}
  \label{fig:fit_P1724}
\end{figure}

The {\em ROSAT} HRI observation of P1724 comprises 13 satellite orbits (see
Fig.~\ref{fig:fit_P1724}). Similar to the flare on SR\,13, constant
count rate is observed only during the very first orbit. We, therefore, base
our value for the quiescent emission, $0.04\,{\rm cps}$, on this time interval 
and find that it is consistent with most of the observations
of P1724 presented by \citey{Neuhaeuser98.1}. 
However, in March 1991 the count rate was higher by a factor 4, possibly 
indicating long-term variations in the quiescent emission.

The lightcurve during the second orbit resembles a small flare event and is
omitted from our analysis.
The maximum of the large flare that dominates this observation extends over 
almost 4 hours. During the decline of the count rate irregular variations are
observed that might be due to short timescale activity superposed on the
large flare event. We ignore these fluctuations and model the lightcurve 
beginning after the second data gap by Eq.~(\ref{eq:fit}).

The total number of source counts measured in this observation is smaller
than 1000, and thus far
too low for a timesliced hardness ratio analysis. 
Having in view the similarity between the 
X-ray lightcurve of this flare and the previously discussed flares, and the
good description of the data by our best fit, we trust that the 
`rotating flare model' applies also to this observation.

\subsection{V773\,Tau}\label{subsect:V773}

An intense X-ray flare on V773\,Tau has been reported by \citey{Skinner97.1} 
and interpreted as a sinusoidal variation
whose period is approximately equal to the known optical period
of V773\,Tau, i.e. 71.2\,h.

\begin{figure}
  \resizebox{\hsize}{!}{\includegraphics{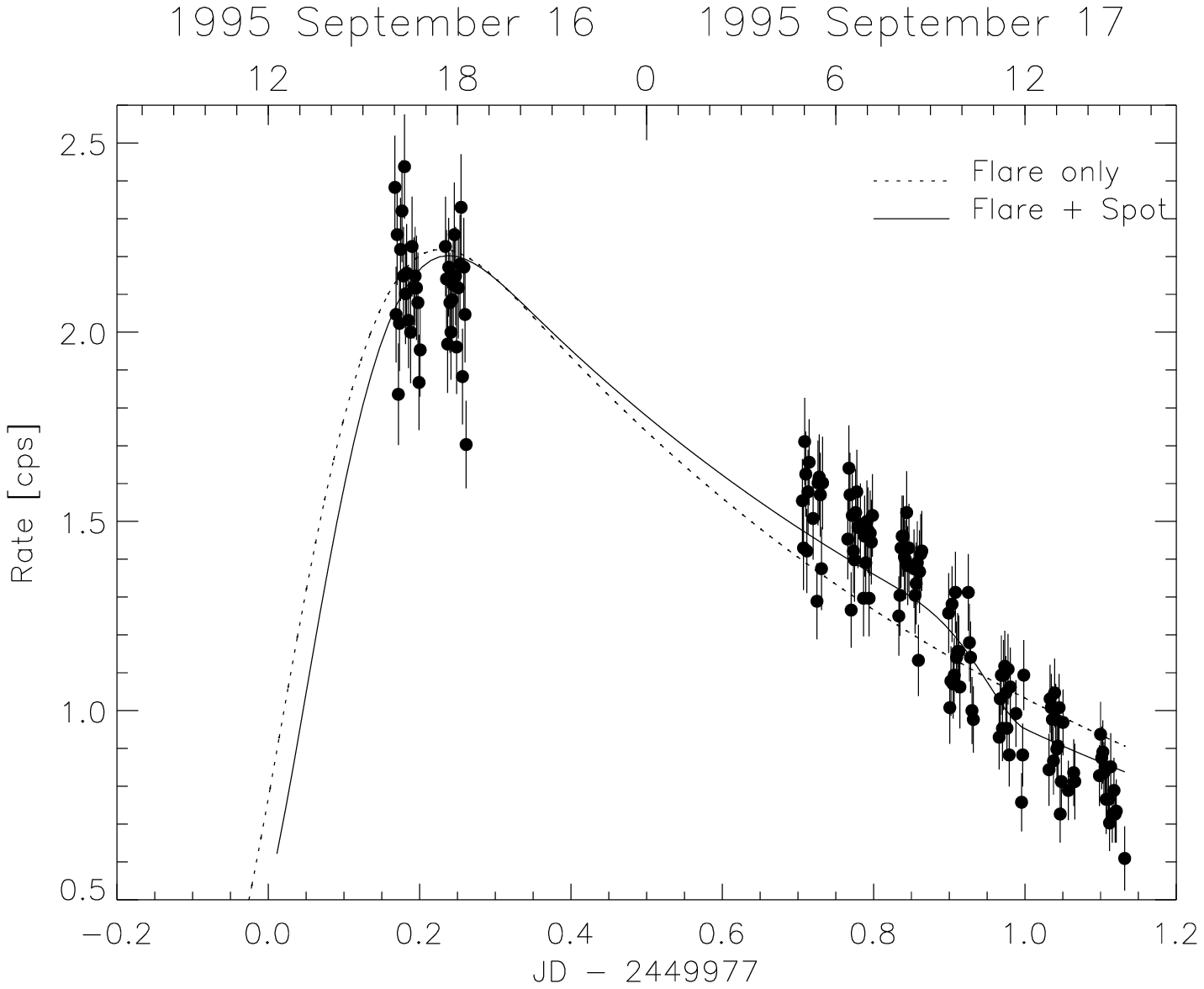}}
  \resizebox{\hsize}{!}{\includegraphics{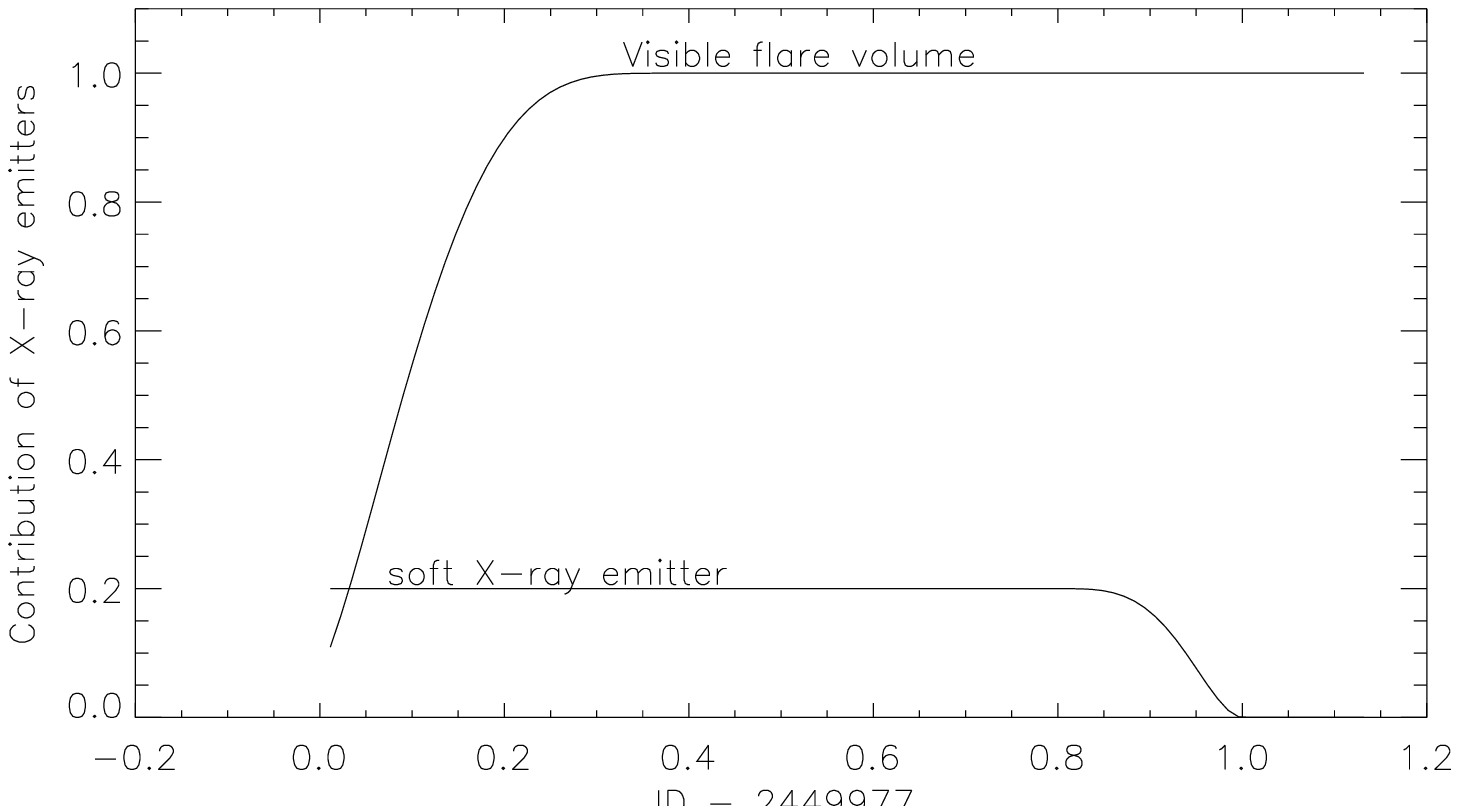}}
  \caption{(a) {\em ASCA} lightcurve of the observation of V773\,Tau (128\,s bins,
  1\,$\sigma$ uncertainties) and best fit. Modeling of a rotating flare
  alone (dotted line) does not lead to an acceptable fit, while adding an 
  additional feature producing enhanced X-rays until JD 2449977.85 and then 
  gradually disappearing describes the bending of the lightcurve and improves 
  the fit perceptibly (solid line).
  (b) Time evolution of the visible flare volume and the secondary X-ray emitter
  during the flare on V773\,Tau as found from the best fit of the
  `rotating-flare model' to the lightcurve.}
  \label{fig:fit_V773Tau}
\end{figure}

The {\em ASCA} lightcurve of this event (see Fig.~\ref{fig:fit_V773Tau}) 
is characterized by constant count rate
at maximum emission which lasts over more than 2\,h making the event a 
candidate for a rotationally modulated flare. No data is available prior to the
peak emission, but observations resumed about 10\,h after the maximum and
display a steady decrease in count rate. 
Since the pre-flare stage and the rise of the flare are completely missing
in the data, the flare volume must have emerged from the backside of
the star well before the start of the observation, and an 
additional time offset parameter $\delta t$ has to be included in the fit 
(analogous to the modeling of the flare on Algol), to determine the time
that elapsed between phase $\pi - \phi_{\rm crit}$ (= emergence of the
flare volume) and the first measurement.

Since the flare covers the complete observation
a value for the quiescent count rate, $I_{\rm q}=0.10 \pm 0.02$\,cps, 
was adopted from a later ASCA SIS0 observation in February 1996, also
presented by \citey{Skinner97.1}.
Despite the fact that the broad maximum of the September 1995 lightcurve 
can be explained by the
loop rotating into the line of sight, no satisfying fit could
be obtained for the flare on V773\,Tau by the model of Eq.~(\ref{eq:fit})
even after a time offset $\delta t$ was added ($\chi^2_{\rm red}=2.51$ for
$149$ degrees of freedom): The decay of the observed 
lightcurve seems to be faster than our model predictions (see 
Fig.~\ref{fig:fit_V773Tau} dotted curve). We note that the data is slightly
bended towards the time axis around the 6th data interval 
after the start of the
observation. This behavior, producing an overall `convex' shape of
the X-ray lightcurve, could be due to an additional feature on the surface
of the star.
We suggest that a localized region with enhanced X-ray emission can
be responsible for this break if this region disappears due to the 
star's rotation
at $\sim {\rm JD}\,2449977.85$. For comparison we show a fit of our
`rotating flare model' where such a feature has been included (solid line in 
Fig.~\ref{fig:fit_V773Tau}, $\chi^2_{\rm red}=1.47$ for $149$ degrees of 
freedom). Since we are interested in a 
qualitative description of the shape of the lightcurve only we 
assumed that this region makes up for
0.2\,cps during its visibility and begins to disappear gradually at JD 2449977.85.
Representing this X-ray emitter by another set of {\em free} parameters would
certainly further improve the already good agreement between data and model. 

\citey{Skinner97.1} 
also present the time behavior of the emission measure
derived from a two-temperature fit to the {\em ASCA} spectrum. 
If our interpretation
of adding a soft X-ray spot, which gradually rotates away, is correct,
then the emission measure of the soft component should stay constant for
most of the time, but decrease towards the end of the observation. However,
the S/N of the time-sliced spectral fits (\cite{Skinner97.1}, his Fig.~10, 
middle panel) is not sufficient to judge whether this is indeed the case.

To conclude, other interpretations such as a different kind of 
anomalous flaring cannot be excluded from the data of this observation. 
\citey{Tsuboi98.1} have presented  
another {\em ASCA} flare observation of V773\,Tau. In that observation 
V773\,Tau shows the typical flare behavior in the sense of a sharp rise and a 
subsequent longer decay of count rate, temperature, and emission measure.
However, their attempt to fit an e-folding decay to the lightcurve of the
hard X-ray count rate was not successful the count rate remaining too
high towards the end of the observation. Hence, unusually long decays
seem to be characteristic for V773\,Tau.

\section{Summary and Conclusions}\label{sect:conclusions}

We have presented a sample of four untypical flare events and provided a
common explanation: Parameters that depend on the size of
the emitting plasma volume (e. g. count rate, $EM$) 
deviate from the standard exponential
decay behavior due to temporary occultation of the flaring volume by the
rotating star. This is most evident from the large flare on Algol,
for which the data are most abundant and our modeling is therefore most
reliable.

The comparatively slow rise of the count rate in the 
X-ray lightcurve, broad maxima and following exponential decay
are well represented by a model that describes emission from a spherical
plasma loop that emerges from the back of the star 
and gradually rotates into the
line of sight of the observer. The increasing visible fraction of the loop
produces the flat maximum and apparent slow-down of the rise stage.

In our data there is no indication for sine-like modulation of the 
X-ray lightcurve, since all but one of the lightcurves are clearly 
asymmetric, and the duration of all events is well below the rotational 
period of the host star. The only exception is the flare on V773\,Tau where 
\citey{Skinner97.1} proposed sinusoidal modulation to reproduce the shape 
of the {\em ASCA} lightcurve. We suggest a different explanation involving
a second X-ray emitting region on the star additionally to a rotationally
modulated flare to come up for the `convex' shape of the lightcurve.
However, due to the lack of pre-flare data, no conclusive evidence is
present for either of the interpretations.
Evidence for reheating of the plasma at $\sim {\rm JD} 2449978.05$ 
inferred from the increase of the hardness ratio 
(see \cite{Skinner97.1}) does not contradict our model, but could be
related to the disappearing of a region emitting {\rm soft} X-rays similar
to the one we introduce.
No significant change in temperature nor emission measure
of the soft component was observed 
in Skinner's spectral analysis, but we note that the results of 
spectral fitting depend on the assumption for the abundances and
column density. 

The decay timescales $\tau$ found from our best fit to the 
respective flare event are all in the typical range
for TTS flares (of a few hours) except for V773\,Tau, where the flare
lasted extraordinary long ($> 20\,{\rm h}$).
Comparatively large loop sizes of a considerable fraction of the radius of
the star are obtained for all observations analysed here, $r$ spanning 
between 10 to 65\,\% of the star radius. 
These values 
are in agreement with typical loop sizes for TTS flares 
inferred from quasi-static loop modeling (see \cite{Montmerle83.1}, 
\cite{Preibisch93.1}).

In view of the large relative size of the ratio between loop and star radius
the assumptions we explain in
Sect.~\ref{sect:model} concerning our model for a rotating flare 
might seem somewhat oversimplifying. We also note that different solutions
of the model seem to describe the data equally well even in the case of the
well restrained Algol observation. Therefore, uncertainties in the fit
parameters are to be regarded carefully.
However, 
the qualitative description of the scenario is very good and the data are
well represented by the model. Other interpretations of the `anomalous'
flare events we presented in this paper may not be excluded but 
are still to be traced.

Clearly, continuous observations of whole flares are needed to verify
whether an event could be subject to rotational modulation of the kind
we discussed in this paper. Up to date most of the flares observed lack
completeness in that either the rise or part of the decay were missed by
the observation. In the near future {\em XMM} will provide the 
possibility of long, uninterrupted observations (up to $\sim 25$\,h)
that will enable to pursue the development of flares in whole. 
Better statistics are needed to be able to study the time development of
spectral parameters for TTS flares, and try to
confirm the `rotating flare model' by use of 
the spectral information similar to our analysis of the Algol observation.

\begin{acknowledgements}

We would like to thank R. Stern and S. Skinner who provided us the {\em
Ginga} and {\em ASCA} data. The {\em ROSAT} project is supported by the 
Max-Planck-Gesellschaft and Germany's
federal government (BMBF/DLR). RN acknowledges grants from the Deutsche
Forschungsgemeinschaft (Schwerpunktprogramm `Physics of star formation').

\end{acknowledgements}

\end{document}